# Epitaxial TiO$_x$ Surface in Ferroelectric BaTiO$_3$: Native Structure and Dynamic Patterning at the Atomic Scale

*Maya Barzilay,*[1,2] *Tian Qiu,*[3] *Andrew M. Rappe*[3] *and Yachin Ivry*[1,2,*]

[1]Department of Materials Science and Engineering, Technion – Israel Institute of Technology, Haifa 3200003, Israel

[2]Solid State Institute, Technion – Israel Institute of Technology, Haifa 3200003, Israel

[3]Department of Chemistry, University of Pennsylvania, Philadelphia, Pennsylvania 19104-6323, U.S.A.

[*]Correspondence to: ivry@technion.ac.il.

**Abstract**

Surfaces and interfaces of ferroelectric oxides exhibit enhanced functionality, and therefore serve as a platform for novel nano and quantum technologies. Experimental and theoretical challenges associated with examining the subtle electro-chemo-mechanical balance at metal-oxide surfaces have hindered the understanding and control of their structure and behavior. Here, we combine advanced electron-microscopy and first-principles thermodynamics methods to reveal the atomic-scale chemical and crystallographic structure of the surface of the seminal ferroelectric BaTiO$_3$. We show that the surface is composed of a native < 2-nm thick TiO$_x$ rock-salt layer in epitaxial registry with the BaTiO$_3$. Using electron-beam irradiation, we successfully patterned artificially TiO$_x$ sites with sub-nanometer resolution, by inducing Ba escape. Therefore, our work offers electro-chemo-mechanical insights into ferroelectric surface behavior in addition to a method for scalable high-resolution beam-induced chemical lithography for selectively driving surface phase transitions, and thereby functionalizing metal-oxide surfaces.

1. **Introduction**:

The ability of oxides to endure extreme mechanical chemical and thermal conditions has been intriguing for researchers from a broad range of disciplines, *e.g.*, earth science, nuclear engineering, space technology and dental care.[1–3] Subtle electro-chemo-mechanical balance at complex-oxide surfaces and interfaces allows the formation of phases with structural and functional characteristics that differ from the bulk.[4–6] The unique electric and magnetic properties of such phases arise from their divergence from stoichiometry, leading to variations in oxidation states of the participating ions. Hence, functional-oxide surfaces and interfaces constitute a rich platform for novel phases that are attractive for high-performance miniaturized electronic devices[7,8] as well as for chemical catalyses.[9]

Because ferroelectric oxides comprise regions with varying crystallographic and electric polarization orientations, there has been a growing interest in their outer surface and domain wall functionality, which is expressed as enhanced conductivity,[10–12] magnetism[4] and even superconductivity.[13] The chemical origin of such functional behavior is typically attributed to either oxygen vacancy dynamics[14–17] or cation segregation,[18] while other studies look at the effects of intrinsic symmetry breaking.[4] Despite the accumulated knowledge on domain walls, the structure and behavior of ferroelectric surfaces, which are responsible for domain stabilization and are attractive for e.g. nano lithography[19–23] and catalysis,[9,24–27] has remained elusive. Specifically, the longstanding challenge in understanding how the surface mediates between the absence of electric and mechanical fields in the vacuum and the polarization, and strain in the bulk is not merely experimental or theoretical, but even conceptual.[28] Hence, computational methods that were developed to explain the electro-chemical[29] and electro-mechanical[30] atomic-scale interactions in ferroelectrics have been adopted to describe experimental observations of the surface behavior. For example, Tsurumi *et al.*[31,32] combined dielectric measurements and DFT calculations to demonstrate that nanoparticles of the seminal non-toxic ferroelectric, $BaTiO_3$ organize in a core-shell structure that helps release strain. These authors suggested that the unit cells at the shell (a few nanometers thick) assume a cubic structure that helps mediate between the tetragonal symmetry of the core and the vacuum. Likewise, a collaborative experimental (STM) – computational (DFT thermodynamics) study demonstrated the stability of various titanium-oxide terminations during surface reconstruction in $BaTiO_3$.[33] By contrast, although transition electron microscopy (TEM) provides us with significant input regarding domain-wall structure and functionality at the atomic scale,[34–40] such TEM characterization of the surface is lacking. Consequently, experimental or theoretical data regarding the surface formation and dynamics or even the surface structure, is absent.

## 2. Results and Discussion:

Combining advanced and conventional electron-microscopy techniques for structural, chemical and oxidation-state analyses with computational methods, we looked at the native chemical, ionic and crystallographic structure of BaTiO$_3$ in ca. 50-nm particles. Moreover, by increasing the dose exposure we used the electron beam to excite the surface, allowing us to image its real-time dynamics. Such particles are on the one hand large enough to be considered bulk-like,[31] while on the other hand they are thin enough to allow atomic-resolution TEM imaging. We demonstrate that the native surface of the tetragonal BaTiO$_3$ crystal is composed of a non-stoichiometric (*i.e.* high defect concentration) nearly cubic titanium oxide phase (TiO$_x$, $x \approx 1$). We also show that by increasing the dose exposure of the material to the electron beam, the electron beam can be used for chemical patterning with sub-nanometer controllability. That is, using the electron beam as a source for exerting localized electric field and heat, we induced Ba escape contactless and expanded the native TiO$_x$ phase by will with nearly atomic resolution on the cost of the bulk BaTiO$_3$. Our imaging methods allowed us not only to realize the Ba escape mechanism, but also to quantify the process.

To reveal the structure near the native surface, we imaged with high-resolution transition electron microscopy (HRTEM) the particles from both {100} and {110} zone axes (ZAs) as seen in Figures 1a-c and 1d-f, respectively. We also performed high-angle annular dark-field imaging (HAADF- STEM) from these ZAs (Figures 1c, f), in which the brightness of each atomic column is proportional to the atomic weight squared ($Z^2$), allowing us to deduce the chemical structure near the surface. The electron micrographs clearly show a difference between the bulk tetragonal BaTiO$_3$ unit cells and a ca. 1-nm thick terrace-like epitaxial structure of different chemical and crystallographic footprints. The HAADF images (Figures 1c, g) indicate that although columns of both Ti and Ba appear clearly in the bulk region (oxygen atoms are too light to be detected in these images), the surface region contains only Ti columns and no Ba. A careful look at the surface in the HRTEM images, where the oxygen columns can be detected reveals neighboring columns with different contrast only from {110} ZA (Fig. 1e-h) and not from the {100} ZA (Fig. 1a-d). Combining the information from the HAADF-STEM and HRTEM methods suggests that the columns are composed of alternating Ti-O columns that can be observed separately only from {110} ZA. Likewise, the inter-atomic distance is similar for both ZAs (eliminating the possibility of rutile or anatase structures). Therefore, we can conclude confidently that the native surface of the BaTiO$_3$ is a rock-salt-like titanium oxide. This conclusion is in agreement with our measurement of the Ti-Ti inter-atomic distance at the interface (2.09±0.005 Å, see Fig. 1b, f), which is substantially smaller than in BaTiO$_3$ (3.99-4.04 Å[41]) and is in agreement with the 4.18 Å lattice parameter of TiO[42] (we should note that this value is much smaller than *e.g.* the 5.54 Å lattice parameter of BaO).

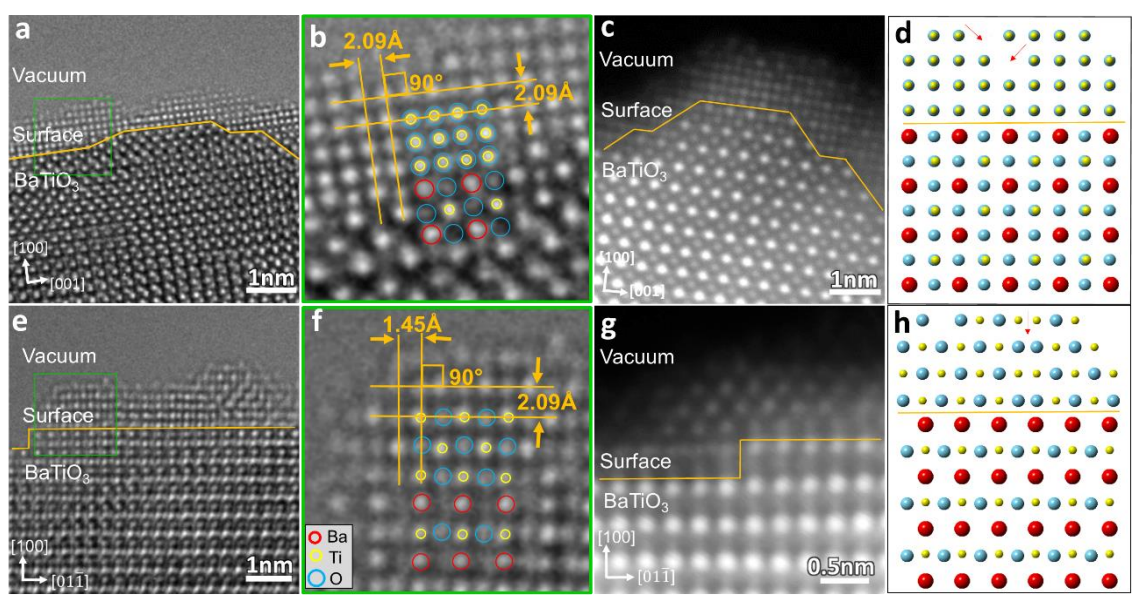

**Figure 1| Epitaxial TiO surface layer in BaTiO$_3$ crystals**. (**a**) High-resolution TEM of the BaTiO$_3$ surface from a [010] zone axis. (**b**) A closer look at the area highlighted in (a) shows the rock-salt structure of the TiO surface as well as the epitaxial growth of the TiO surface on the perovskite BaTiO$_3$ crystal. (**c**) HAADF image of a different grain from a similar zone axis. (**d**) Schematic illustration of the [100] non-stoichiometric TiO$_x$ surface – BaTiO$_3$ bulk native structure. The complementary (**e**) large-scale and (**f**) closer-look high-resolution TEM images as well as (**g**) HAADF image taken from the [110] zone axis shows the BaTiO$_3$ crystal and rock-salt TiO from a perpendicular orientation. (**h**) Schematic illustration of the [110] non-stoichiometric TiO$_x$ surface – BaTiO$_3$ bulk native structure. The border between the bulk BaTiO$_3$ and the TiO surface is highlighted (orange lines in a, c, e and g), while the location of Ba, Ti and O atoms is designated in (b) and (f).

To support our chemical analysis, we used a novel microscopy system that allows us to perform atomic-resolution electron-diffraction x-ray spectroscopy (EDX) mapping[43] simultaneously with HAADF imaging at the same region. Figure 2 shows such cross-identification of the Ba, Ti and oxygen columns. Likewise, to further support our structural analysis, we used integrated differential phase contrast (iDPC)[44] imaging, which is a method that has recently been developed for mapping the location of individual atomic columns with very high accuracy (also simultaneously with the HAADF and EDX imaging). In this method, the contrast is proportional to the atomic weight ($\propto Z$), allowing for high detectability of heavy and light atoms alike (see the **Experimental Section** for more details about the iDPC and EDX imaging). Figure 2h shows the iDPC signal from a small area near the surface (same area as in Fig. 1c), verifying the exact location of the rock-salt-like $TiO_x$ atomic columns (as well as of the perovskite $BaTiO_3$ structure).

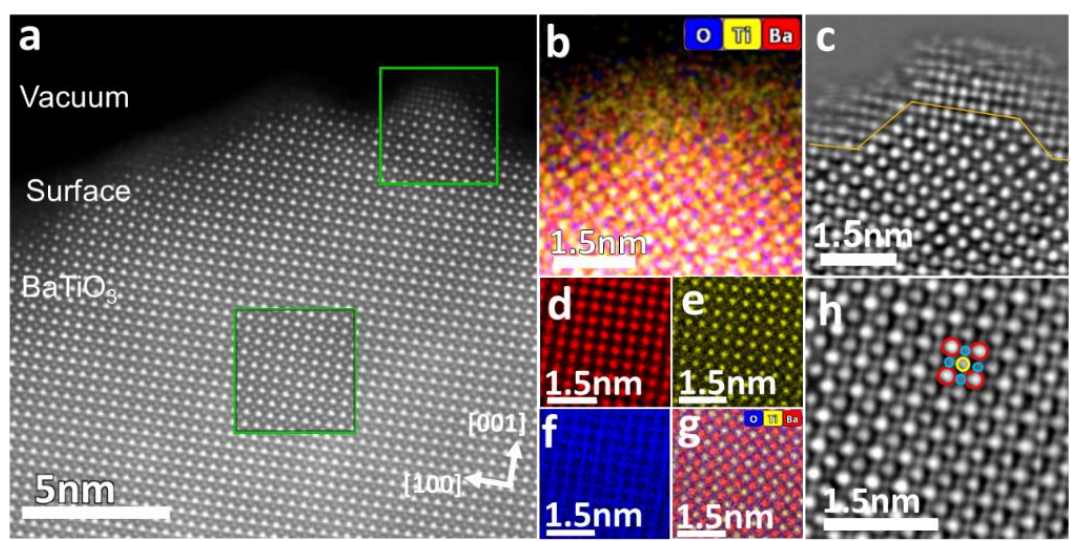

**Figure 2| Cross-chemical and structural analysis of the BaTiO₃ crystal and TiO surface**. (**a**) HAADF image of the BaTiO₃ crystal and its surface. (**b**) Atomic-scale chemical mapping (EDX) showing higher Ti concentration and lower Ba concentration at the surface (here, the signal is not of an instantaneous beam detection. Rather, the signal is integrating over a large time, hence deteriorating the imaging stability and the spatial resolution). (**c**) High-accuracy location mapping (iDPC) of the same area demonstrates the oxygen-titanium rock-salt structure of the TiO surface. Simultaneous high-resolution chemical analysis of the (**d**) barium, (**e**) titanium and (**f**) oxygen EDX signals as well as the combined chemical analysis (**g**) and high-precision location mapping (iDPC) are given as a reference. The areas in (a) from which the surface and bulk images were taken are highlighted, while the HAADF image of the same surface area is given in Figure 1c.

We should note that the universality of the existence of a native TiO$_x$ layer at the surface of BaTiO$_3$ was confirmed by a large-area panoramic view (Supporting Information, Figure 3), while the appearance of a TiO$_x$ surface was observed in all the different BaTiO$_3$ crystals we examined (> 50 different crystals), including the five different particles that are presented in this paper. These crystals arrived from different sources and were prepared by different methods (see **Experimental Section**). Moreover, revisiting existing literature (*e.g.* Figure 5 from Zhu *et al*.[45]), we believe that the TiO$_x$ surface has been observed previously, but the chemical and crystallographic structure have not yet been identified probably due to the lack of the novel imaging methods that have been used in the current study. Finally, the thickness of the native TiO$_x$ surface layer is ca. 1 nm suggests that there are only one to three monolayers, and hence this observation is in agreement with the previous observations of titanium-oxide termination in BaTiO$_3$,[33] which typically cannot determine the exact structure of the inner layers.

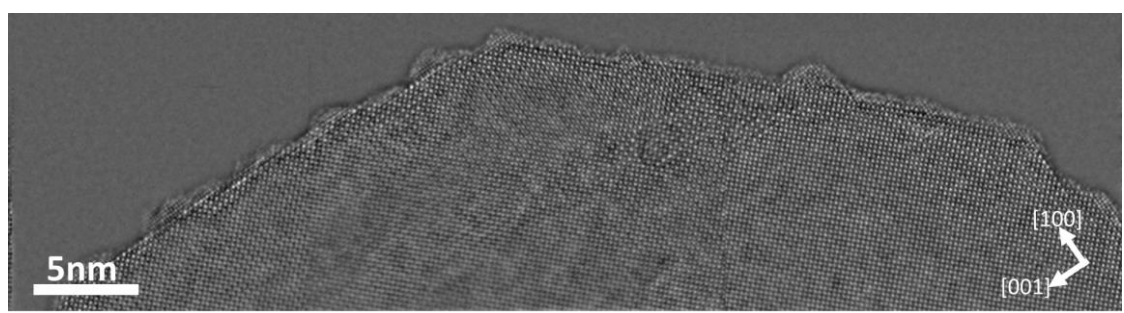

**Figure 3| Long-range presence of the native epitaxial TiO$_x$ surface of BaTiO3 crystals.** High-precision atomic-scale mapping (iDPC) of the native TiO$_x$ surface over a long range. The panoramic image is composed of four iDPC images.

The deviation from TiO stoichiometry in a rock-salt structure (TiO$_x$ with $1.3 > x > 0.7$) leads to high defect concentration.[46] Using the 3D nature of TEM imaging, we identified three types of defects at the very thin native surface: stacking fault, twinning and vacancies. These defects appear clearly in Figure 4. Moreover, using electron energy-loss spectroscopy (EELS) signals, we were able to identify the chemical shift[47] of the Ti peaks between the ions at the bulk and the ions at the surface. This shift corresponds to the variations in the Ti oxidation state, while no shift was observed in either the Ba or O peaks. Figure 5 shows that the position of the Ti cation peaks recorded from the bulk BaTiO$_3$ corresponds to Ti$^{4+}$. However, the signal recorded from the surface demonstrates that the peak shifts towards lower energies, indicating on the existence of Ti$^{2+}$, as expected from the TEM imaging.[47–49]

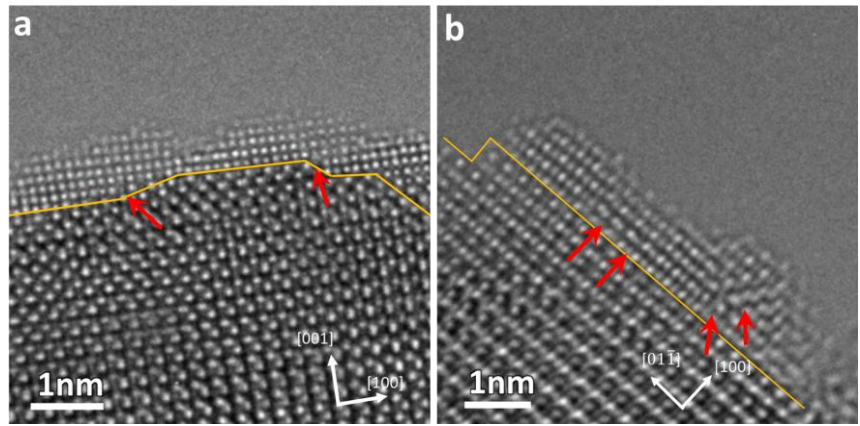

**Figure 4| Defects in the native TiO$_x$ surface.** Highlighted (red arrows) crystallographic defects in the TiO surface at the areas presented (**a**) in Figure 1a and (**b**) in Figure 1d.

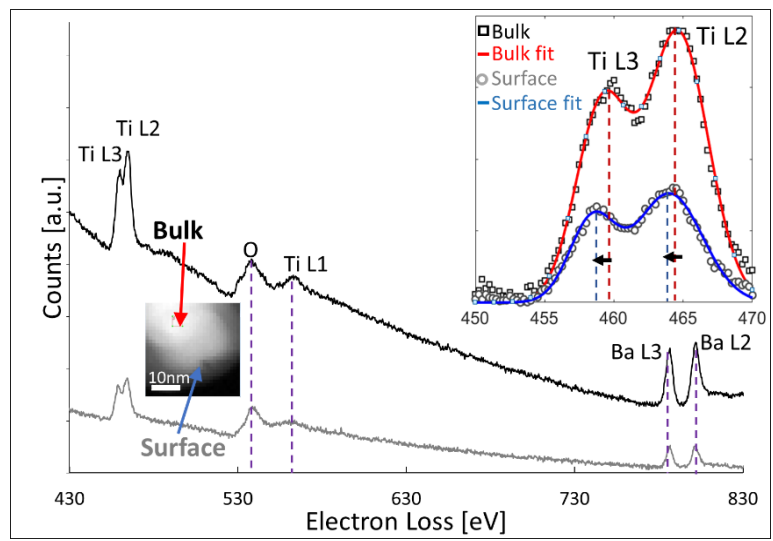

**Figure 5| Chemical shift between the bulk and surface Ti binding energies in BaTiO$_3$.** A comparison between the EELS signals from the bulk and from the surface (HAADF image of the areas from which the EELS signals were taken is given as a reference). Although no change was observed in the picks related to the Ba or to the O ions, a closer look (insert) at the L2 and L3 energies of the Ti ions show a chemical shift that corresponds to the reduction in oxidation state from Ti$^{4+}$ in the perovskite BaTiO$_3$ bulk towards the Ti$^{2+}$ in rock-salt structure TiO surface.[47–49]

The experimental identification of a TiO epitaxial surface for native BaTiO$_3$ indicates that TiO is a stable phase on the BaTiO$_3$ surface. Because the stability of heterogeneous thin films depends strongly on the interfacial interactions and cannot be directly predicted from their bulk stability, we perform *ab initio* DFT calculations to determine the stability of different surfaces on BaTiO$_3$ and to derive the stability map of this interface. To cover a large span of possible surface phases, we include all phases in the Ba-Ti-O$_2$ ternary system found in Materials Project[50] that can be epitaxially placed on BaTiO$_3$ (001) surface. The energy of each phase P$_i$ is expressed in Equation (1).

$$E_{P_i} = E_{BTO+P_i} - E_{BTO} \tag{1}$$

Where E$_{BTO+P_i}$ is the total energy of BaTiO$_3$ with P$_i$ phases on the surface, and E$_{BTO}$ is the energy of the bare BaTiO$_3$ crystal. A phase is considered stable if none of the decomposition paths is spontaneous, *i.e.*, when for any other three phases P$_j$, P$_k$, and P$_l$ that satisfy Equation (2).

$$P_i \rightarrow n_j P_j + n_k P_k + n_l P_l \tag{2}$$

where n$_j$, n$_k$, and n$_l$ are stoichiometric numbers, the following inequality holds in Equation (3):

$$E_{P_i} \leq n_j E_{P_j} + n_k E_{P_k} + n_l E_{P_l} \tag{3}$$

where energies are calculated from Equation (1). Similarly, two phases P$_i$ and P$_j$ could coexist if, for any other two phases P$_k$ and P$_l$, as given in Equation (4)

$$P_i + n_j P_j \rightarrow n_k P_k + n_l P_l \tag{4}$$

P$_i$ and P$_j$ have lower energy, as shown in Equation (5)

$$E_{P_i} + n_j E_{P_j} \leq n_k E_{P_k} + n_l E_{P_l} \tag{5}$$

Figure 6 shows that based on this method (more computational details can be found in Supporting Information), we not only show the stability of the native TiO$_x$ layer, but we also predict that a thicker layer may also be stable, while the stoichiometry (*x*) varies with the film thickness. This effect can be seen in the difference between the ternary diagrams of two atomic monolayers (Figure 6a) and four atomic monolayers (Figure 6c) of TiO$_x$ on the BaTiO$_3$ surface. We think that introducing external excitations in the experimental conditions discussed here (*i.e.* TEM) a removal of oxygen from the system can be induced, driving it down from the point of BaTiO$_3$ into a ternary region on the phase diagram. This ternary region evolves from BaTiO$_3$-BaO-Ti$_4$O$_5$ to BaTiO$_3$-BaO-TiO when increasing thickness of overlayers, suggesting that $Ti_4O_5$ is the stable composition for the thinner $TiO_x$ layer (Figure 6Figure 6Figure 6a-b), while TiO is also stable for thicker layers (Figure 6c-d).

These results allow us to draw two main conclusions regarding the observed surface overlayers. First, TiO incurs less lattice mismatch penalty compared with $Ti_4O_5$ because the relative stability of TiO increases for thicker TiO layers. This conclusion is confirmed by comparing the lattice mismatch from their bulk states, where $Ti_4O_5$ has 5.0% lattice mismatch and TiO has 3.5% lattice mismatch. Secondly, $Ti_4O_5$ has a stronger interaction with the $BaTiO_3$ (001) lattice plane. A consequence is that the $TiO_x$ layer closest to the $BaTiO_3$ (*e.g.* the native $TiO_x$) is $Ti_4O_5$, while subsequent layers of TiO may be induced. These results are in close agreement with our experimental measurements, while our prediction of the $TiO$-$Ti_4O_5$ interplay may also explain the observations of high defect concentration (Figure 4). Finally, these calculations allow us to predict that introducing external excitations to the $BaTiO_3$-$TiO_x$ system may favor controlled growth of the pseudo-cubic $TiO_x$ ($Ti_4O_5$-TiO) near the native $TiO_x$ surface. We should note that Ti is smaller than Ba and hence presumably we expect that the growth of the surface will include barium oxide formation rather than titanium oxide formation. However, Ba ions doe not have many intermediate oxidation states and tend to be stable only as $Ba^{2+}$. Hence, as soon as the surface becomes suboxide barium oxide would become unstable. On the other hand, Ti has a broader range of oxidation-state stability and can even be stable as $Ti^{3+}$, e.g. in $Ti_2O_3$. This analysis is in agreement also with previous studies regarding $BaTiO_3$ surface termination that show that under reducing conditions (i.e., when oxygen atoms are driven away), the Ba also goes away, leaving a Ti-rich level.[33]

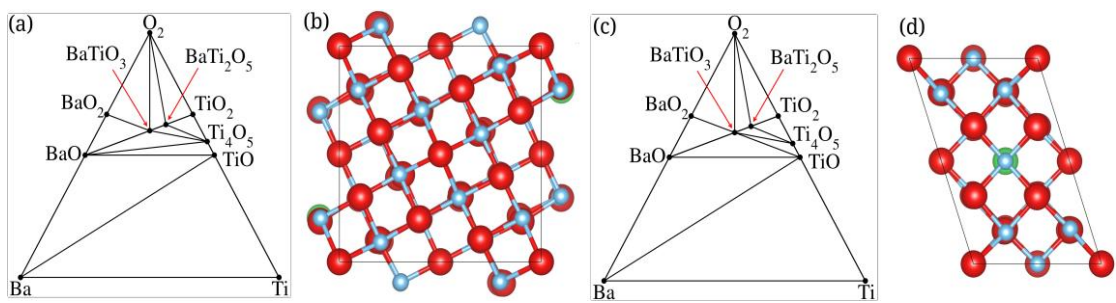

**Figure 6| Calculated phase-stability map of the BaTiO3 surface.** (a) Phase diagram of phases with thickness of two layers on $BaTiO_3$ surface. (b) Top view of $Ti_4O_5$ on $BaTiO_3$ surface. (c) Phase diagram of phases with thickness of four layers on $BaTiO_3$ surface, noticing that the ternary region $BaTiO_3$-BaO-TiO replaces the ternary region $BaTiO_3$-BaO-$Ti_4O_5$ observed in the two-layer case. (d) Top view of TiO on $BaTiO_3$ surface.

Following the prediction that exciting the system can grow the TiO$_x$ phase, we used the electron-beam irradiation as a contactless source for such excitations at a length scale that is determined by the beam size. Because we knew that TEM imaging does not affect the surface, the attempts to excite the surface included increase of the electron-beam doses with respect to the doses that are used for TEM imaging. Figure 7a-d shows the growth of the TiO$_x$ surface on the expense of BaTiO$_3$ as a function of electron-beam irradiation time over a large area. Calculating carefully (see **Experimental Section** and Fig. SI2) the irradiation of the high-resolution TEM electron beam, the dose was 0.003 nA nm$^{-2}$. The growth of the TiO$_x$ surface as a function of dose exposure is clearly observed at the images. We should note that for very long exposure times, we can confidently assume that the TiO$_x$ grew significantly also at the direction parallel to the beam and not only perpendicular to the surface, so that a distinction between the bulk BaTiO$_3$ and the TiO$_x$ surface is not so clear any more (see Figure 7d). Thanks to the accurate EDX measurements we were able to quantify the rate of Ba escape (Figure 7f-g), *i.e.* the dynamic BaTiO$_3$-to-TiO$_x$ transition under a constant electron-beam irradiation. Here, the dose was much higher that in the TEM imaging: 80 nA nm$^{-2}$ at 200 keV (see **Experimental Section** and Fig. SI2 for further details regarding the dose calculation), so that the Ba escape process was also much faster. Given the above theoretical analysis regarding the Ba escape, we can deduce that the electron-beam irradiation lowers the oxidation state of the metal atoms, so that to become more stable, there must be less oxygen atoms around them. This state of oxygen deficiency would favorite Ba escape over Ti escape, even though the Ba atoms are larger.

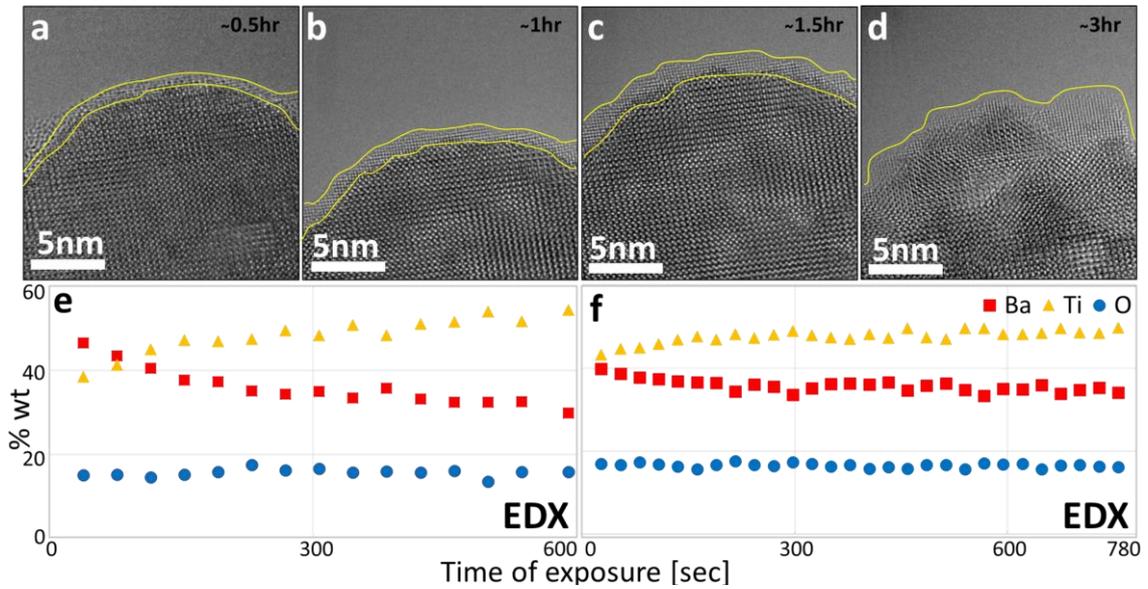

**Figure 7| Dynamic growth of epitaxial TiO$_x$ surface in BaTiO$_3$.** High-resolution TEM images of a surface area irradiated with high doses of an electron beam (0.003 nA nm$^{-2}$) after ca. (**a**) 30 min; (**b**) 60 min; and (**c**) 90 min show the dynamic evolution of the TiO surface. (**d**) Long-time irradiation helps grow the TiO surface not only from the side but also from the top and bottom surfaces of the BaTiO$_3$. (the BaTiO$_3$-TiO$_x$ border is therefore more difficult to trace with TEM, where the imaging averages the integrated signal of electrons that travel through the entire sample- top surface, bulk and bottom surface). (**e**) Schematic illustration of the chemical patterning of the non-stoichiometric TiO$_x$ surface of the ferroelectric BaTiO$_3$. (**f**) EDX chemical analysis of the surface evolution (same area as in Figure 2b), showing the TiO growth and Ba escape as a function of time. iDPC of the native epitaxial TiO layer (before $t = 0$ sec) and the TiO that was grown due to the irradiation-induced Ba escape (after $t = 590$ sec) are given in Fig. 8, while the dynamic formation of the TiO layer during the Ba escape as recorded by the HAADF imaging is given in Supporting Information Video S1. (**g**) EDX analysis of the TiO layer grown over a thicker area of the particle (bulk) under 80 nA nm$^{-2}$ electron flux irradiation (same area as in Figure 2d-g).

We then wanted to examine whether such chemical patterning can be performed also at higher-resolution, using a smaller beam size. Supporting Information Video S1 and Figure 8 show real-time TiO$_x$ patterning in sub-nanometer resolution as imaged with iDPC (made simultaneously with the EDX characterization). Here, the higher currents accelerate the formation of the TiO$_x$ layer as expected. The dynamics of the high-resolution TiO$_x$ patterning shows that the process involves intermediate formation of an amorphous state (see Supporting Information, Video S1 and Figure 8), while the ability to tune the beam parameters with high accuracy (energy and spatial resolution) allows us to control the TiO$_x$ patterning.

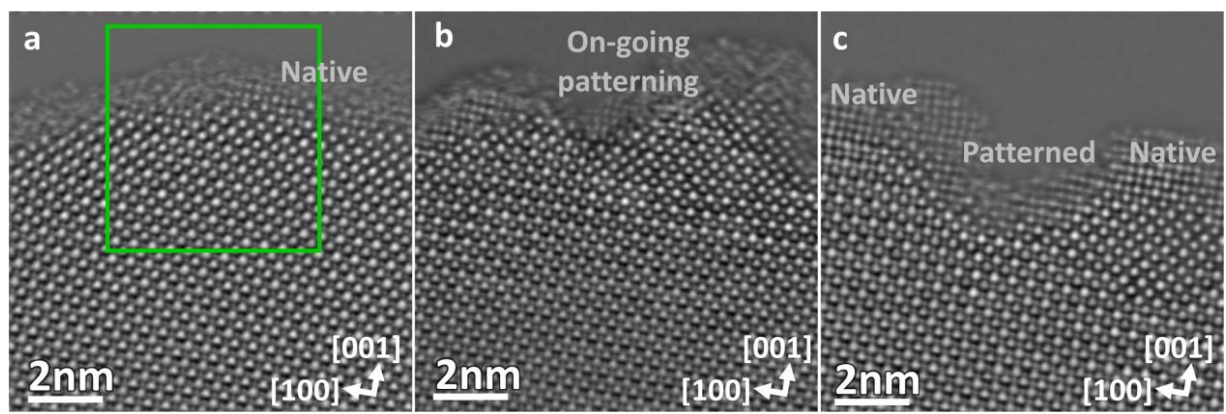

**Figure 8| High-precision atomic-scale location mapping of the BaTiO$_3$ surface during intentional TiO$_x$ growth.** iDPC images of the bulk BaTiO3 structure and TiO surface (**a**) before the exposure to intense electron beam for EDX data collection (before $t = 0$ sec in Figure 6e) .(**b**) Three minutes after the exposure (after $t = 600$ sec in Figure 6e). After the Ba escaped during the irradiation, the formed TiO layer crystallizes slowly so that it is not yet completely an epitaxial crystalline layer. (**c**) Thirty minutes after the exposure. The TiO is fully crystallized. Note that the area was not exposed to the electron beam after the irradiation caused by the EDX data collection, only for a brief moment during the readout of (b) and (c). The dynamic formation of the TiO layer (before crystallization) as a function of dose irradiation was imaged in HAADF mode as given in Supporting Information Video S1. The area that was exposed to high doses, from which the video was taken is designated in a green square in image (a).

We should note that the dose window for TiO$_x$ patterning is rather narrow as on the one hand, the energy should be higher than a certain threshold value that induces Ba escape. On the other hand, larger doses are too violent and might also evaporate the Ti and O ions from the surface, *i.e.* etch away the entire BaTiO$_3$ (electron-beam irradiation of $\gtrsim$270 nA nm$^{-2}$ at 200 keV even for short periods of time, see Supporting Information Video S2). We should also note that as opposed to the quantitative EDX data (Figures 7e and f), the EDX imaging is an integration of an accumulated signal, allowing us to increase the signal-to-noise ratio above an imaging-relevant threshold. Hence, EDX imaging of the surface, where the material is rather thin and produces even weaker signals, requires large integration time. Given the fluctuation in STEM (thermal, electromagnetic etc.), the long integration time gives rise to reduction in the spatial resolution. Such a decrease in the EDX imaging resolution is seen e.g. in the case of Figure 2b. Moreover, if the exposure time is too long, the beam might also induce chemical and structural changes (e.g. amorphization and Ba escape), reducing the imaging resolution even further. Yet, the quantitative data (e.g. Figure 7e-f) can be evaluated accurately even as a function of time, because it is based on spatial integration rather than on time integration.

## 3. Conclusion

Our results can explain previous observations of a Ti-rich surface in $BaTiO_3$ crystals,[51] while the observed dynamic growth of the TiO layer under the electron beam supports earlier predictions of Ba escape from the $BaTiO_3$ surface,[33, 45] giving rise to enhanced surface reactivity.[52] The existence of cation-escape mechanisms near the surface, including as a consequence of electron-beam irradiation[53] as well as the effects of these mechanisms on enhanced surface reactivity has been reported recently in other ferroelectrics.[54] However, the exact phases of these surfaces have not been predicted or determined experimentally. Hence, we encourage future examination of the generality of the $TiO_x$ surface formation in other Ti-based ferroelectric oxides. It is possible that the presence of the cubic $TiO_x$ (or $TiO_x$) at the surface is in agreement with the hypothesis of Tsurumi *et al.*, who suggest that the surface of the tetragonal ferroelectric must be cubic, for stabilizing the polar matter-air discontinuity. Nevertheless, we see that the cubic surface is not the $BaTiO_3$ ferroelectric, but rather it is a chemically distinct phase, TiO, which can also be grown by will.

Similarly to mechanical mediation between the bulk and the vacuum, we would like to suggest a possible electric mediation between the ferroelectric crystal and the vacuum. Although the observed high defect concentration may be explained partially be the co-existence of TiO and $Ti_4O_5$, we would like to raise another potential explanation. Being a ferroelectric, the ions in the $BaTiO_3$ organize in dipole moments that give rise to the ferroelectric polarization domains. These polarization domains (that are switchable) should somehow be stabilized at the $BaTiO_3$ surface. The $TiO_x$ thin surface layer takes place between the air or the vacuum and the ferroelectric $BaTiO_3$. Thus, the $TiO_x$ may serve as a dielectric mediator, which helps compensate for the built-in ferroelectric polarization in the $BaTiO_3$, and hence effectively screens the polarization. $TiO_x$ is typically metallic around $x \approx 1$, and thus might not be considered as a dielectric medium in the first instance. However, although we cannot measure the $TiO_x$ surface resistance directly, we can assume that the metallic properties are suppressed significantly in ultra-thin films as can be deduced e.g. from the temperature dependence of electrical resistance of ultra-thin cubic TiO films.[55] Thus, we would like to propose that the defects in the $TiO_x$ may help modulate the electric field locally for supporting the local polarization distribution along the surface of the $BaTiO_3$. Alternatively, the ability to pattern locally integrated epitaxial metallic $TiO_x$ electrodes may be helpful for integrated ferroelectric electronics. In either case, the existence of off-stoichiometry metal oxide surface may be used for redox-based chemical catalysis. It is also interesting to note that thin epitaxial layers of cubic TiO have been characterized as superconductors with $T_C$ up to 7.4 K.[55–58] Thus, the coverage of the BaTiO$_3$ crystals with a functional TiO$_x$ surface and the ability to pattern this phase locally may help nanoscale fabrication of superconducting-based technologies. Following the growing interest in electron-beam patterning,[59] we believe that it will be interesting to test whether this method of chemical patterning of surfaces by using localized electron beam may be applied also to other materials, with an emphasize on ternary metal-oxide phases, for enhancing their surface functionality.

4. **Experimental Section**:

We used commercially available BaTiO$_3$ nanoparticle (99.9% purity spherical particles, purchased from US Research Nanomaterials, Inc.) as well as particles from collaboration with Prof. Yoed Tsur- (Technion not presented in this work). The NPs were suspended in ethanol and sprayed on an amorphous Carbon-Cu grid by introducing high pressure N$_2$, allowing the individual NPs spread evenly on the grid.

All in-situ irradiation experiments and imaging in this work was done using double corrected Titan Themis G2 300 (FEI / Thermo Fisher) HRTEM with sub-angstrom resolution. both TEM and STEM images were taken with 300-200 keV acceleration voltage, with medium-low monochromated current of 2-0.1nA depending on using HRTEM or HRSTEM mood. No manipulation has been conducted to the images.

In order to investigate the surface of BTO NPs, different TEM methods were used. Structural analysis was performed by high-resolution (HR) observations. Atomic-scale chemical mapping was done by Dual-X detector: Energy-dispersive X-ray spectroscopy (EDX) detector, drift correction was applied. Accurate positions of the heavy elements and *Z* contrast analysis were performed using high-angle annular dark-field (HAADF). Accurate position and *Z* contrast analysis of the light elements were performed using 4 quadrants detectors for the novel HRSTEM Integrated Differential Phase Contrast (iDPC-STEM) technique.

Flux calculations were done as follows. For the HRTEM: we divided the last measured screen current by the beam size (the beam is steady and larger them the CCD camera, with 30-50 nm diameter). For HRSTEM (or iDPC): we divided the last measured screen current by the pixel size (the beam scans the sample and its diameter is smaller than the pixel size: 0.04-0.02 nm with 512x512 pixels for EDX/HAADF mapping and 0.008-0.004 nm with 2048x2048 pixels for iDPC/HAADF images). See Figure SI2 for further details. We should note that during the initial alignment of the system for finding the desired zone axis, the particles were exposed to some irradiation. Based on Figure 7Figure 7, we estimate this exposure to be negligible, in particular for HRSTEM modes, were the area examined was mostly different than the area used for the alignment. Yet, in some case, minor contribution of the beam to enhance the native layer is possible.

The advantage of using simultaneously HAADF and iDPC is the complementary information they provide. The HAADF is very sensitive to atomic mass and less affected by non-uniform thickness in the sample. However, the contrast in the HAADF images is proportional to $Z^2$. Therefore, when imaging together light with heavy elements by HAADF, the light elements are usually invisible.

The iDPC- STEM technique is sensitive to the electrostatic potential of the atoms, and the contrast in images is relative to Z. Using the iDPC technique drastically improves the detectability of light elements among heavy elements in the same image.

Computational and modeling: for detailed explanations regarding the methods used for the modeling and computational analysis presented in this work please refer to the relevant Section in the SI.

**Supporting Information (SI)**

Supporting Information is available from the Wiley Online Library or from the author.

**Acknowledgements**

The Technion group acknowledges financial support from the Zuckerman STEM Leadership Program, the Horev Fellowship for Leadership in Science and Technology, supported by the Taub Foundation and the Russel Barry Nanoscience Institute, from the Eliyahu Pen Research Fund as well as from the Israel Science Foundation (ISF) grant # 1602/1. Likewise, we would like to thank Dr. Yaron Kaufman and Mr. Michael Kalina for technical support, while we thank Prof. Yoed Tsur for supplying us with some of the $BaTiO_3$ crystals. The theory and modeling research (of TQ and AMR) was supported by the U. S. Department of Energy, Office of Science, Office of Basic Energy Sciences, under Award DE-SC0019281. The authors thank the NERSC of the U. S. Department of Energy for computational support.

**Authors' contribution**

MB conducted the microscopy imaging and measurements as well as analyzed the data; TQ performed the quantum mechanical and thermodynamic surface calculations; AMR initiated, supervised and helped analyze the computational work; YI initiated and supervised the research, helped analyze the data and wrote the paper. All authors took an active part in preparing the manuscript.

Gebauer, U. Gerstmann, C. Gougoussis, A. Kokalj, M. Lazzeri, L. Martin-Samos, N. Marzari, F. Mauri, R. Mazzarello, S. Paolini, A. Pasquarello, L. Paulatto, C. Sbraccia, S. Scandolo, G. Sclauzero, A. P. Seitsonen, A. Smogunov, P. Umari, R. M. Wentzcovitch, *J. Phys. Condens. Matter* **2009**, *21*, 395502.

[63] A. M. Rappe, K. M. Rabe, E. Kaxiras, J. D. Joannopoulos, *Phys. Rev. B* **1990**, *41*, 1227.

[64] N. J. Ramer, A. M. Rappe, *Phys. Rev. B* **1999**, *59*, 12471.

[65] http://opium.sourceforge.net.

**Supporting Information (SI)**

**Phases included in DFT calculation**

In the case of epitaxial growth, the material should have a lattice plane that can match the BaTiO$_3$ (001) surface with small distortion, Figure SI1 a-b show the matched lattice plane of Ba$_2$Ti$_6$O$_{13}$ and Ti$_3$O$_5$, respectively. Based on this condition, we set the tolerance to be 10% and found 21 candidates from the full set of 154 materials in the Ba-Ti-O$_2$ ternary system in the Materials Project [50] database. These materials with Materials Project IDs are listed in Table (1).

Table 1: Table of all phases involved in DFT calculation

| Composition | Materials Project ID | Composition | Materials Project ID |
| --- | --- | --- | --- |
| Ba$_2$Ti$_3$O$_7$ | mvc-1214 | Ba$_2$Ti$_3$O$_8$ | mvc-14876 |
| Ba$_2$Ti$_6$O$_{13}$ | mp-7733 | Ba | mp-10679 |
| BaO$_2$ | mp-1006878 | BaO$_2$ | mp-1105 |
| BaO | mp-1342 | BaTi$_2$O$_5$ | mp-3943 |
| BaTi$_2$O$_5$ | mp-555966 | Ti$_3$O$_4$ | mp-755875 |
| Ti$_3$O$_5$ | mp-1147 | Ti$_4$O$_5$ | mp-10734 |
| Ti | mp-6985 | Ti | mp-73 |
| TiO$_2$ | mp-25262 | TiO$_2$ | mp-390 |
| TiO$_2$ | mp-554278 | TiO$_2$ | mvc-11912 |
| TiO | mp-1203 | TiO | mp-2664 |
| TiO | mp-755300 | | |

**Details of DFT calculation**

We put the relevant phases on top of four layers of BaTiO$_3$ in the simulation cell. 15Å vacuum is added in the *z* direction with a dipole correction[60] to cancel the artificial interaction between the system and its periodic images. The cell parameters in the *xy* plane are calculated from the BaTiO$_3$ tetragonal bulk phase. During structural relaxation, the bottom two BaTiO$_3$ layers are fixed and the rest of the layers in the cell are allowed to relax. A sketch of the supercell is shown in Figure SI1c. For all DFT calculations, we used the Perdew-Burke-Ernzerhof functional revised for solids (PBEsol)[61] as implemented in the QUANTUM ESPRESSO package.[62] All atoms are represented by norm-conserving, optimized,[63] designed nonlocal[64] pseudopotentials generated with the OPIUM package,[65] treating the 5*s*, 5*p*, 5*d*, and 6*s* of Ba, 3*s*, 3*p*, 3*d*, and 4*s* of Ti, 2*s* and 2*p* of O as semi-core and valence states.

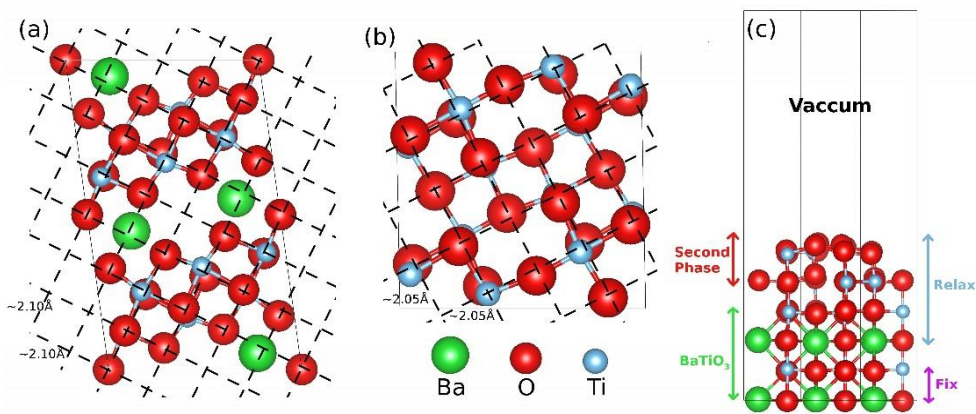

**Figure SI1| The DFT calculated phases and supercell.** (**a**) Lattice match in case of Ba$_2$Ti$_6$O$_{13}$. (**b**) Lattice match in case of Ti$_3$O$_5$ (**c**) Sketch of the supercell in DFT calculations.

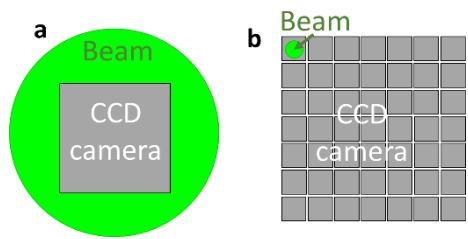

**Figure SI2| schematic illustration of the electron flux calculation.** (**a**) HRTEM mode, were the electron beam is larger than the CCD camera, (**b**) HRSTEM mode, were the electron beam is smaller than each pixel.

**Video S1| Dynamic patterning of TiO at the BaTiO$_3$ surface (LINK)**. HAADF images of (100) BaTiO$_3$ surface during high-intensity electron-beam irradiation of 80 nA nm$^{-2}$ at 200 keV for 590 sec (i.e. the EDX data collection that was presented in Figure 6e). The video shows the evolution of the TiO layer during Ba escape and the formation of amorphous TiO layer before recrystallization. Recrystallization occurred at the area that was intentionally irradiated, but when the beam was not directed at the area (and hence this process could have not been imaged).

**Video S2| Violent evaporation of BaTiO$_3$ from the surface (LINK)**. HAADF images of (110) BaTiO$_3$ surface during high-intensity electron-beam irradiation of 270 nA nm$^{-2}$ at 200 keV for 392 sec, showing a very fast amorphization that is associated with the Ba escape, followed by rapid evaporation of the Ti and O atoms from the BaTiO$_3$ surface, hence challenging the controllability of the TiO patterning.